\documentstyle[preprint,aps,prd]{revtex}

\def\lprox{\mathrel{\raise .3ex\hbox{$<$\kern-
.75em\lower1ex\hbox{$\sim$}}}}

\def\gprox{\mathrel{\raise .3ex\hbox{$>$\kern-
.75em\lower1ex\hbox{$\sim$}}}}

\def\be{\begin{equation}}
\def\ee{\end{equation}}
\def\ben{\begin{eqnarray}}
\def\een{\end{eqnarray}}

\begin{document}

\draft

\title{\bf  Decay of massive scalar hair in the background\\ of a black hole with a global mononpole }

\author{Hongwei Yu}

\address{Institute of Physics, Hunan Normal University,\\
    Changsha,  Hunan 410081, China}

%\date{}

\maketitle

\tightenlines

\begin{abstract}
The late-time tail behaviors of massive scalar fields are examined
analytically in the background of a black hole with a global
monopole. It is found that the presence of a solid deficit angle
in the background metric makes the massive scalar fields decay
faster in the intermediate times. However, the asymptotically
late-time tail is not affected and it has the same decay rate of
$t^{-5/6}$ as in the Schwarzschild  and nearly extreme
Reissner-Nordstr\"{o}m backgrounds.
\end{abstract}


\section{introduction}
Ever since Wheeler put forward the  no-hair theorem \cite{W},
which states that the external field of a black hole relaxes to a
Kerr-Newman type characterized solely by the black-hole's  mass,
charge and angular momentum, there have been a lot of
investigations concerning the dynamical mechanism by which
perturbations fields outside a black hole are radiated away. The
massless scalar, gravitational and electromagnetic external
perturbations were first studied by Price \cite{Price} in the
Schwarzschild background and an inverse power-law tail,
$t^{2l+3}$,  has been found to dominate the late-time behavior of
these perturbations for a fixed position, if there is no initial
static field. Here $l$ is the multiple moment of the wave mode and
$t$ is the Schwarzschild time coordinate. The late-time behaviors
of these massless neutral perturbations along the null infinity
and along the future event horizon were further examined by
Gundlach et al \cite{GPP1,GPP2}. Recently the late-time tail has
also been considered in the case of a rotating black hole by
Barack and Ori \cite{BaO}.

Although these works are mainly concerned with massless fields,
the evolution of massive scalar fields is also important and it
has attracted a lot of attention recently.  Behaviors
qualitatively different from those of massless fields have been
found. For instance, It has been shown in Ref.\cite{HandP} that an
oscillatory power-law tail of the form $ \sim t^{-l-{3\over
2}}\sin (\mu t)$ for massive scalar fields develops at the
intermediate late-time characterized by $ \mu M\ll 1$ in
Reissner-Nordstr\"{o}m background. Here $\mu $ is the mass of the
scalar field and $M$ is that of the black hole.  Note that the
massive scalar fields decay slower than massless ones. It should
be pointed out, however, that this intermediate tail is not the
final pattern that dominates at very late times \cite{HandP}. In
fact, a transition from the intermediate behavior to  an
oscillatory tail with the decay rate of $t^{-5/6}$ has been
demonstrated to occur at asymptotically late times both in the
Schwarzschild and nearly extreme Reissner-Nordstr\"{o}m
backgrounds \cite{KandT,KandT1}.

In this paper, we will examine both the intermediate and
asymptotic late time behaviors for massive scalar fields at a
fixed radius in the background of a black hole with a global
monopole. A global monopole is one of the topological defects that
may have been formed  during phase transitions in the evolution of
the early universe and a black hole with a global monopole is the
result of an interesting process in which a black hole swallows a
global monopole \cite{BV}.  An unusual property of the
black-hole-global-monopole system is that it possesses a solid
deficit angle, which makes it quite different topologically from
that of a Schwarzschild black hole alone. The physical properties
of the black-hole-global-monopole system have been studied
extensively in recent years. These include,  but are not limited
to, the gravitational \cite{BV,HL} and the vacuum polarization
effects \cite{ML,MBK}, the particle creation in the formation of
the system \cite{L}, the black hole thermodynamics \cite{HY1}, and
more recently the energy spectra of non-relativistic quantum
system in the background\cite{MB}.

Our purpose here is to see what effects the solid deficit angle in
the background metric due to the presence of a global monopole
will have on the late-time evolution of massive scalar fields. In
Sec. II we describe the physical system and formulate the problem
in terms of the black-hole Green's function using the spectral
decomposition method \cite{Leaver}. In Sec. III, we examine both
the intermediate and asymptotic late-time behaviors of massive
scalar fields. We conclude in Sec. IV with a brief summary and
some discussions.

\section{ description of the system and Green's function formalism}

 We examine the time evolution of a massive scalar field in the
background of a black hole with a global monopole. The metric is
described by
\begin{equation}
ds^2= -\biggl( 1-8\pi G\eta_0^2- {2Gm\over r}
\biggr)\;dt^2+\biggl( 1-8\pi G\eta_0^2- {2Gm\over r}
\biggr)^{-1}\;dr^2+ r^2d\Omega^2 \;, \label{eq:metric1}
\end{equation}
where $ m $ is the mass of the black hole and $\eta_0$ is the
symmetry breaking scale when the monopole is formed \cite{BV} .
Introducing the coordinate transformation
\begin{equation}
t\longrightarrow(1-8\pi G\eta_0^2)^{\frac12}t, \qquad
r\longrightarrow (1-8\pi G\eta_0^2)^{\frac{-1}2}r
\end{equation}
and new parameters
\begin{equation} M=(1-8\pi G\eta_0^2)^{-3/2}m, \qquad
b=1-8\pi G\eta_0^2
\end{equation}
then we can rewrite metric Eq.~(\ref{eq:metric1}) as
\begin{equation}
ds^2=-\left(1-\frac{2GM}r\right)dt^2+\left(1-\frac{2GM}r\right)^{-1}dr^2
+r^2 b(d\theta^2+\sin^2\theta d\phi^2)\label{eq:metric2}
\end{equation}
This metric is, apart from the deficit solid angle $\Delta=4\pi
b=32\pi G\eta_0^2$, very similar to the Schwarzschild metric and
we will use this form thereafter. The equation of motion for a
minimally coupled scalar field with mass $\mu$ is
\begin{equation}
  \label{eq:K-G}
  \frac{1}{\sqrt{-g}}\frac{\partial}{\partial
x^{\mu}}\bigg(\sqrt{-g}g^{\mu\nu}\frac{\partial\phi}{\partial
x^{\nu}}\biggr) -\mu^2\phi=0\;.
\end{equation}
$\phi$ can be separated in the given metric as
\begin{equation}
  \label{eq:harmonics}
  \phi =\sum _{l,m}\frac{\psi ^l(r)}{r}Y_{lm}(\theta ,\varphi),
\end{equation}
hereafter we omit the index $l$ of $\psi ^l$ for simplicity.
Using the tortoise coordinate $r_{\ast} $ defined by
\begin{equation}
  \label{eq:tortoise}
  dr_{\ast}=\frac{dr}{1-\frac{2M}{r}}\>,
\end{equation}
 we obtain a wave equation for each multiple moment:
\begin{equation}
  \psi _{,tt}-\psi _{,r_{\ast}r_{\ast}}+V\psi =0,
\end{equation}
where the effective potential $V$ is
\begin{equation}
  V=\left(1-\frac{2M}{r}\right)
\left[\frac{l(l+1)b^{-1}}{r^2}+\frac{2M}{r^3} +\mu^2\right].
\end{equation}
Define the retarded Green's function $G(r_{\ast},r_{\ast}';t)$ by
\begin{equation}
  \label{eq:retarded}
  \left[\frac{\partial ^2}{\partial t^2}
-\frac{\partial ^2}{\partial r_{\ast}^2} +V
\right]G(r_{\ast},r_{\ast}';t) = \delta (t)\delta
(r_{\ast}-r_{\ast}')\;,
\end{equation}
for $t>0$.  The causality condition gives the initial condition
that $G(r_{\ast},r_{\ast}';t)=0$ for $t\le 0$. Then the time
evolution of the massive scalar field is given by
\begin{equation}
  \psi (r_{\ast},t)=\int \left[G(r_{\ast},r_{\ast}';t)\psi _t(r',0)
+ G_t(r_{\ast},r_{\ast}';t)\psi (r_{\ast}',0) \right] dr_{\ast}'
\end{equation}
In order to find $G(r_{\ast},r_{\ast}';t)$ we use the Fourier
transform
\begin{equation}
  \label{eq:fourier}
  \tilde{G}(r_{\ast},r'_{\ast};\omega)
=\int _{0^{-}}^{+\infty} G(r_{\ast},r'_{\ast};t)e^{i\omega t}dt.
\end{equation}
The Fourier transform is analytic in the upper half $\omega$
plane, and the corresponding inversion formula is
\begin{equation}
  \label{eq:inverse}
  G(r_{\ast},r_{\ast}';t)= -\frac{1}{2\pi}\int _{-\infty +ic}^{\infty +ic}
\tilde{G}(r_{\ast},r'_{\ast};\omega) e^{\scriptscriptstyle
-i\omega t}d\omega
\end{equation}
where $c$ is some positive constant.

Let $\tilde\psi_1(r_{\ast},\omega)$ and
$\tilde\psi_2(r_{\ast},\omega)$ be two linearly independent
solutions to the homogenous equation
\begin{equation}
  \label{eq:homo}
  \left(\frac{d^2}{dr_{\ast}^2}+\omega ^2 -V\right)
\tilde{\psi}_i =0 \quad i=1,2.
\end{equation}
The Green's function can be constructed as follows
\begin{equation}
  \label{eq:Green}
 \tilde{G}(r_{\ast},r'_{\ast};\omega )= -\frac{1}{W(\omega)}
\left\{
\begin{array}{l@{\quad,\quad}l}
\tilde{\psi} _1(r'_{\ast},\omega) \tilde{\psi
}_2(r_{\ast},\omega)&
\qquad  r'_{\ast} >r _{\ast},\\
\tilde{\psi} _1(r_{\ast},\omega) \tilde{\psi }_2(r'_{\ast},\omega)
&\qquad r'_{\ast}< r_{\ast} .
\end{array}
\right.
\end{equation}
Here $W(\omega)$ is the Wronskian defined as
\begin{equation}
W(\omega )=\tilde{\psi}_1\tilde{\psi}_{2,r_{\ast}}
-\tilde{\psi}_{1,r_{\ast}}\tilde{\psi}_2.
\end{equation}
To calculate $ G(r_{\ast},r_{\ast}';t)$ using
Eq.~(\ref{eq:inverse}), one needs to close the contour of
integration into the lower half of the complex frequency plane. It
has been argued that the asymptotic tail is associated with the
existence of a branch cut (in $\tilde\psi_2$) placed along the
interval $-m \le \omega \le m$ \cite{Leaver,HandP}. This tail
arises from the integral of $\tilde{G}(r_{\ast},r'_{\ast};\omega
)$ around the branch cut (denoted by $G^C$) which gives rise to an
oscillatory inverse power-law behavior of the field. Therefore our
goal is to evaluate $G^C (r_{\ast},r'_{\ast};\omega )$.

\section{Evolution of massive scalar fields}

Now let us assume that both the observer and the initial data are
situated far away from the black-hole such that $r \gg M$. We
expand the wave-equation (\ref{eq:homo}) in $M/r$ to obtain
(neglecting terms of order $O[({{M} \over r})^{2}]$ and higher)
\begin{equation}\label{eq:field}
  \left[ {{d^2} \over {dr^2}} +w^{2}-\mu^{2} +{{4Mw^{2}-2M\mu^{2}} \over r} -
{{l(l+1)b^{-1}} \over {r^{2}}} \right ] \xi =0\  ,
\end{equation}
where $\xi=(1-{2M\over r})^{1/2} \tilde \psi$. This equation can
be solved in terms of Whittaker's functions. The two basic
solutions needed to construct the Green's function are (for $|w|
\leq m$)
\begin{equation}\label{sol1}
\tilde \psi_1 =M_{\kappa,\rho}(2\varpi r)\  ,
\end{equation}
and
\begin{equation}\label{sol2}
\tilde \psi_2 =W_{\kappa,\rho}(2\varpi r)\  ,
\end{equation}
where
\begin{equation}
\varpi=\sqrt{\mu^2-\omega^2},\quad\quad \kappa={3\over
2}{M\mu^2\over \varpi}-2M\varpi,  \quad\quad \rho=\sqrt {
l(l+1)b^{-1}+{1\over 4}}.
\end{equation}
Let us note that these solutions can also be written in terms of
two standard confluent hypergeometric functions, $ M(a,b,z)$ and
$U(a,b,z)$, as follows,
\begin{equation}
\tilde \psi_1 =M_{\kappa,\rho}(2\varpi r)=e^{-\varpi r}(2\varpi
r)^{{1\over 2}+\rho}\;M(\rho+{1\over2}-\kappa,\;1+2\rho,\; 2\varpi
r)\ ,
\end{equation}
and
\begin{equation}
\tilde \psi_2 =W_{\kappa,\rho}(2\varpi r)=e^{-\varpi r}(2\varpi
r)^{{1\over 2}+\rho}\;U(\rho+{1\over2}-\kappa,\;1+2\rho,\;
2\varpi r) \  ,
\end{equation}

Using Eq. (\ref{eq:inverse}), one finds that the branch cut
contribution to the Green's function is given by
\begin{eqnarray}\label{eq:GC}
G^C(r'_{\ast},r_{\ast};t)&=&{1 \over {2\pi}} \int_{-\mu}^{\mu}
\left[{\tilde \psi_1(r_{\ast}',\varpi e^{\pi i})  {\tilde
\psi_2(r_{\ast},\varpi e^{\pi i})} \over {W(\varpi e^{\pi i})}} -
{\tilde \psi_1(r_{\ast}',\varpi ){\tilde \psi_2(r_{\ast},\varpi
)} \over
{W(\varpi )}} \right] e^{-iwt} dw\nonumber\\
&=&{1 \over {2\pi}} \int_{-\mu}^{\mu} f(\varpi)e^{-iwt} dw\;.
\end{eqnarray}
For simplicity we assume that the initial data has a considerable
support only for $r$-values which are smaller than the observer's
location. This, of course, does not change the late-time behavior.
Let us note that when $t$ is large, the term $e^{-iwt}$ oscillates
rapidly. This leads to a mutual cancellation between the positive
and the negative parts of the integrand, so that the effective
contribution to the integral arises from $|w|$=$O(\mu-{1 \over
t})$ or equivalently $\varpi = O(\sqrt {{\mu \over t}})$
\cite{HandP}.

Using the following relations
\begin{eqnarray}
W_{\kappa,\;\rho}(2\varpi r)&=&{\Gamma(-2\rho)\over
\Gamma({1\over2}-\rho-\kappa)}\;M_{\kappa,\;\rho}(2\varpi r)
 \nonumber\\&&+{\Gamma(2\rho)\over
\Gamma({1\over2}+\rho-\kappa)}\; M_{\kappa,\;-\rho}(2\varpi r)\;,
\end{eqnarray}
and
\begin{equation}
M_{\kappa,\; \rho}(e^{\pi i}2\varpi r)=e^{({1\over2}+\rho)\pi
i}M_{-\kappa, \;\rho}(2\varpi r)\;,
\end{equation}
we find, with the help of 13.1.20 of Ref. \cite{Abramo}, that
\begin{equation}
W(\varpi e^{\pi i})=-W(\varpi)={\Gamma(2\rho)\over
\Gamma({1\over2}+\rho-\kappa)}\;4\rho\varpi\;,
\end{equation}
and consequently,
\begin{eqnarray}
f(\varpi)&=&{1\over4\rho\varpi}\;\left[ M_{\kappa,\;\rho}(2\varpi
r'_{\ast})M_{\kappa,\;-\rho}(2\varpi
r_{\ast})-M_{-\kappa,\;\rho}(2\varpi
r'_{\ast})M_{-\kappa,\;-\rho}(2\varpi
r_{\ast})\right]\nonumber\\&&+ {1\over4\rho\varpi}
{\Gamma(-2\rho)\Gamma({1\over2}+\rho-\kappa)\over
\Gamma(2\rho)\Gamma({1\over2}-\rho-\kappa)}\biggl[
M_{\kappa,\;\rho}(2\varpi r'_{\ast})M_{\kappa,\;\rho}(2\varpi
r_{\ast}) \nonumber\\&& \quad +e^{(2\rho+1)\pi
i}\;M_{-\kappa,\;\rho}(2\varpi
r'_{\ast})M_{-\kappa,\;\rho}(2\varpi r_{\ast})\biggr]\;.
\end{eqnarray}

\subsection{Intermediate late-time Tails}

First we discuss the intermediate asymptotic behavior of the
massive scalar field. That is the tail in the range
\begin{eqnarray}
  M \ll r\ll t \ll \frac{M}{(\mu M)^2}\;.
\end{eqnarray}
In this time scale, the frequency range $\varpi = O(\sqrt {{\mu
\over t}})$ , which gives the dominant contribution to the
integral, implies
\begin{equation}
\label{eq:inter} \kappa\ll 1\;.
\end{equation}
Notice that $\kappa$ originates from the $1/r$ term in the massive
scalar field equation. It describes the effect of backscattering
off the spacetime curvature.  If the relation Eq.~(\ref{eq:inter})
is satisfied, the backscattering off the curvature from
asymptotically far regions (which dominates the tails of massless
fields) is negligible. However, it is worthwhile to point out that
the intermediate time tail here will be different from that in the
case without a global monopole because of the nontrivial topology
in the background metric , i.e. , the presence of the solid
deficit angle since $b\neq 1$.  So, we have in this case,
\begin{equation}
f(\varpi)\approx {(1+e^{(2\rho+1)\pi i})\over4\rho\varpi}
{\Gamma(-2\rho)\Gamma({1\over2}+\rho)\over
\Gamma(2\rho)\Gamma({1\over2}-\rho)} M_{0,\;\rho}(2\varpi
r'_{\ast})M_{0,\;\rho}(2\varpi r_{\ast})\;.
\end{equation}
Since $\varpi r\ll 1$, the above equation can be further
approximated, by using $ M(a,b,z)\approx 1$ as $z\rightarrow 0$,
to give
\begin{equation}
f(\varpi)\approx {(1+e^{(2\rho+1)\pi
i})\Gamma(-2\rho)\Gamma({1\over2}+\rho)\over4\rho
\Gamma(2\rho)\Gamma({1\over2}-\rho)2^{-2\rho-1}}
(r'_{\ast}r_{\ast})^{{1\over 2}+\rho}\varpi ^{2\rho}\;.
\end{equation}
Substituting the above result into Eq.~(\ref{eq:GC}), we obtain
\begin{eqnarray}
G^C(r'_{\ast},r_{\ast};t)&=& {(1+e^{(2\rho+1)\pi
i})\Gamma(-2\rho)\Gamma({1\over2}+\rho)\over 8\pi\rho
\Gamma(2\rho)\Gamma({1\over2}-\rho)2^{-2\rho-1}}
(r'_{\ast}r_{\ast})^{{1\over
2}+\rho}\;\int_{-\mu}^{\mu}\;\varpi ^{2\rho}\;e^{-iwt}\nonumber\\
&=& {(1+e^{(2\rho+1)\pi
i})\Gamma(-2\rho)\Gamma({1\over2}+\rho)\Gamma(\rho+1)\mu^{\rho+{1\over2}}\over
\sqrt{\pi}\rho
\Gamma(2\rho)\Gamma({1\over2}-\rho)2^{-3\rho-{3\over2}}}
(r'_{\ast}r_{\ast})^{{1\over
2}+\rho}\;t^{-\rho-{1\over2}}J_{\rho+{1\over2}}(\mu t)\;,
\end{eqnarray}
where $J_{\rho+{1\over2}}$ is the Bessel function.  In the limit
$t\gg \mu^{-1}$, it becomes
\begin{equation}
\label{eq:intermediate}
 G^C(r'_{\ast},r_{\ast};t)={(1+e^{(2\rho+1)\pi
i})\Gamma(-2\rho)\Gamma({1\over2}+\rho)\Gamma(\rho+1)\mu^{\rho}\over
\pi\rho \Gamma(2\rho)\Gamma({1\over2}-\rho)2^{-3\rho-2}}
(r'_{\ast}r_{\ast})^{{1\over 2}+\rho}\;t^{-\rho-1}\cos(\mu
t-(\rho+1)\pi/2)\;,
\end{equation}
which clearly exhibits an oscillatory inverse power-law behavior.
Let's note that in general $ b< 1$ and recall that
\begin{equation}
\rho=\sqrt { l(l+1)b^{-1}+{1\over 4}}\nonumber\;,
\end{equation}
then a  comparison of the result here with Eq.~(32) of Ref.
\cite{HandP} tells us that in the intermediate times the power-law
tail depends not only on the multiple number of the wave mode but
also on the the parameter ($b$) characterizing the space-time
metric, and  the massive scalar field decays faster in the black
hole background with a global monopole than in that without it.
So, although the intermediate tail is not affected significantly
by the curvature, it is by the topology of the background metric.

\subsection{asymptotic late-time tails}

In the above calculation, we have used the approximation of
$\kappa\ll 1$, which only holds when $ \mu t \ll 1/\mu^2M^2 $.
Therefore, the power-law tail found in the last section is not the
final one, and a change to a different pattern of decay is
expected when $\kappa$ is not negligibly small. In this section,
we examine the asymptotic tail behavior at very late times such
that
\begin{equation}
\mu t \gg  {1\over \mu^2M^2}.
\end{equation}
Now we have
\begin{equation}
\kappa \simeq {3M\mu^2\over 2\varpi}\gg 1\;.
\end{equation}
So the backscattering off the curvature will be important in this
case. Using Eq.~(13.5.13) of Ref.\cite{Abramo}, we have, in the
limit $\kappa\gg1$, that
\begin{equation}
M_{\pm\kappa,\;\pm \rho}(2\varpi r)\approx \Gamma(1\pm
2\rho)(2\varpi r)^{1\over 2}(\pm \kappa )^{\mp\rho} J_{\pm
2\rho}(\sqrt{\pm \alpha r})\;\,
\end{equation}
where $\alpha=8\kappa\varpi\approx 12M\mu^2$. Consequently, we
have
\begin{eqnarray}
f(\varpi)&\approx
&{\Gamma(1+2\rho)\Gamma(1-2\rho)r_{\ast}'r_{\ast}\over2\rho}\;\left[J_{
2\rho}(\sqrt{\alpha r_{\ast}'\;})J_{ -2\rho}(\sqrt{\alpha
r_{\ast}})-I_{ 2\rho}(\sqrt{\alpha r_{\ast}'\;})I_{
-2\rho}(\sqrt{\alpha r_{\ast}})\right] \nonumber\\&&+
{1\over2\rho}
{\Gamma(1+2\rho)^2\Gamma(-2\rho)\Gamma({1\over2}+\rho-\kappa)r_{\ast}'r_{\ast}\over
\Gamma(2\rho)\Gamma({1\over2}-\rho-\kappa)}\;\kappa^{-2\rho}\;\biggl[J_{
2\rho}(\sqrt{\alpha r_{\ast}'\;})J_{ 2\rho}(\sqrt{\alpha
r_{\ast}}) \nonumber\\&& \quad +I_{ 2\rho}(\sqrt{\alpha
r_{\ast}'\;})I_{ 2\rho}(\sqrt{\alpha r_{\ast}})\biggr]\;,
\end{eqnarray}
where $ I_{ \pm 2\rho}$ is the modified Bessel functions. Clearly,
the late time tail arising from the first term will be $\sim
t^{-1}$. Now let us try to figure out what behavior the second
term gives rise to. For this purpose, we define
\begin{eqnarray}
A&=&{1\over2\rho}
{\Gamma(1+2\rho)^2\Gamma(-2\rho)r_{\ast}'r_{\ast}\over
\Gamma(2\rho)}\;\biggl[J_{ 2\rho}(\sqrt{\alpha r_{\ast}'\;})J_{
2\rho}(\sqrt{\alpha r_{\ast}})+I_{ 2\rho}(\sqrt{\alpha
r_{\ast}'\;})I_{ 2\rho}(\sqrt{\alpha r_{\ast}})\biggr]\;,
\end{eqnarray}
then the contribution from the second term to the Green's function
can be written as
\begin{equation}
{A \over 2\pi}\;\int_{-\mu}^{\mu}\;
{\Gamma({1\over2}+\rho-\kappa)\over
\Gamma({1\over2}-\rho-\kappa)}\;\kappa^{-2\rho}\;e^{-iwt}\;,
\end{equation}
which  can be approximated, in the limit $\kappa\gg1$, by using
\begin{equation}
\Gamma(z)\Gamma(-z)=-{\pi\over z \sin\pi z}\;,
\end{equation}
and Eq.~(6.1.39) in Ref.\cite{Abramo}, as follows
\begin{equation}
\label{eq:Integral}
 {A \over 2\pi}\;\int_{-\mu}^{\mu}\;
\;e^{i(2\pi\kappa-wt)}\;e^{i\phi}\;dw\;.
\end{equation}
Here the phase $\phi $ is defined by
\begin{equation}
e^{i\phi}={ 1+(-1)^{2\rho}\;e^{-i2\pi\kappa} \over
1+(-1)^{2\rho}\;e^{i2\pi\kappa}}\;.
\end{equation}
 The integral Eq.~(\ref{eq:Integral}) is very similar to that of Eq.~(61) in Ref.\cite{KandT} and it  can be evaluated by method of the
 saddle-point integration as in such. Hence the asymptotic late time tail arising from the second term is $ \sim t^{-{5 \over 6}}
 $, and it dominates over the tail from the first term. So, we
 have
\begin{equation}
G^C(r'_{\ast},r_{\ast};t)\sim t^{-{5 \over 6}}\;.
\end{equation}

\section{Summary}
We have studied analytically both the intermediate and
asymptotically late-time evolution of massive scalar fields in the
background of a black hole with a global monopole. We find that if
$\mu M\ll 1$ the intermediate tails given by
Eq.~(\ref{eq:intermediate}) dominates at the intermediate
late-time $\mu M\ll \mu t\ll 1/(\mu M)^2$ at a fixed radius.
Because of the presence of the solid deficit angle in the
background metric, the decay is faster than  those in the
Schwarzschild and Reissner-Nordstr\"{o}m backgrounds
\cite{HandP,KandT,KandT1}. Therefore,  in the intermediate
late-times,  the oscillatory power-law tail depends not only on
the multiple number of the wave mode but also on the the parameter
($b$) characterizing the space-time metric. Hence  although the
intermediate late-time tail is not affected significantly by the
curvature, it is by the topology of the background metric.
 However, the intermediate late-time tail is not the final pattern
 and a transition to an oscillatory tail with the decay rate of
 $t^{-5/6}$ is to occur when $\mu t\gg 1/(\mu M)^2$. The origin of the tail may be attributed
 to the resonance backscattering off the space-time curvature.  It is interesting to note that  this tail behavior  is
 independent of the field mass, the multiple moment of the wave
 mode and the space-time parameter $b$ and it is same as that in
 the black hole backgrounds without global monopoles studied in
 \cite{KandT,KandT1}.  It should be pointed out, however, that this late time tail begins to dominate only when
 $\mu t\gg 1/(\mu M)^2$  i.e. $\kappa\gg 1$. So, the tail of massive scalar
 fields will still be dependent on the multiple moment of the wave mode and the topology  of
 the space-time during the transitional intermediate times when this condition is not
 satisfied.  Our result seems to suggest that the
 oscillatory $t^{-5/6}$ tail may be  a quite general feature for
 the late-time decay of massive scalar fields in any static black hole
 backgrounds.

\acknowledgments

This work was supported by the National Science Foundation of
China under Grant 10075019.

\end{document}